\documentclass[preprint,12pt]{aastex}
\usepackage{epsfig}
\usepackage{emulateapj5}
\usepackage{apjfonts}

\newcommand{\chandra}{{\it Chandra}}

\newcommand{\rosat}{{\it ROSAT}}

\newcommand{\xmm}{{\it XMM-Newton}}

\newcommand{\lum}{\thinspace\hbox{$\hbox{ergs}\thinspace\hbox{s}^{-1}$}}

\newcommand{\sss}{CXOU\,J140332.3+542103}

\makeatletter
\newenvironment{inlinetable}{%
\def\@captype{table}%
\noindent\begin{minipage}{0.999\linewidth}\begin{center}\footnotesize}
{\end{center}\end{minipage}\smallskip}

\newenvironment{inlinefigure}{%
\def\@captype{figure}%
\noindent\begin{minipage}{0.999\linewidth}\begin{center}}
{\end{center}\end{minipage}\smallskip}

\begin{document}

\def\spose#1{\hbox to 0pt{#1\hss}}
\def\laeq{\mathrel{\spose{\lower 3pt\hbox{$\mathchar"218$}}
     \raise 2.0pt\hbox{$\mathchar"13C$}}}
\def\gaeq{\mathrel{\spose{\lower 3pt\hbox{$\mathchar"218$}}
     \raise 2.0pt\hbox{$\mathchar"13E$}}}

\slugcomment{Accepted for publication in ApJL}

\title{Evidence for an Intermediate-Mass Black Hole: \chandra\ and
  \xmm\ Observations of the Ultraluminous Supersoft X-ray Source in
  M101 during its 2004 Outburst}
\author{A.~K.~H.~Kong$^1$, R.~Di\,Stefano$^{1,2}$, F.~Yuan$^3$}
\affil{$^1$ Harvard-Smithsonian Center for Astrophysics, 60
Garden Street, Cambridge, MA 02138; akong@cfa.harvard.edu}
\affil{$^2$ Department of Physics and Astronomy, Tufts
University, Medford, MA 02155}
\affil{$^3$ Department of Physics, Purdue University, 1396 Physics
  Building, West Lafayette, IN 47907}

\begin{abstract}
We report the results of \chandra\ and \xmm\ observations of a new 
outburst of an ultraluminous supersoft X-ray source in M101.
\sss\ was observed in a low luminosity state ($L_X\sim
10^{37}$\lum) between 2004 January and May. The low-state X-ray spectra were
relatively hard; the combined low state spectrum can be fitted with
a combination of a power-law with photon index of 1.4 and a blackbody
of 63 eV. During 2004 July, the
source underwent a strong outburst and the peak 0.3--7 keV luminosity
reached $3\times10^{40}$\lum, with a bolometric luminosity of about
$10^{41}$\lum. The outburst spectra were very soft and can generally be fitted with a
blackbody model with temperatures of 50--100 eV. In two of
the observations, absorption edges at 0.33 keV, 0.56 keV, 0.66 keV,
and 0.88 keV were found. An \xmm\ observation was also performed during the
decay of the outburst, and a power-law tail was seen in addition to
the supersoft spectrum. We consider different accretion models; one
involving an intermediate-mass black hole can explain the observations.

\end{abstract}


\section{Introduction}

Recent high resolution X-ray observations reveal that there are large
number of ultraluminous X-ray sources (ULXs) in many nearby
galaxies. ULXs are usually defined as very bright ($L_X \approx
10^{39-41}$\lum) off-nucleus, variable X-ray sources in
galaxies. Since the luminosity exceeds the Eddington luminosity of a
$10M_\odot$ black hole (BH), ULXs are considered to be candidates for
intermediate-mass black holes (IMBHs). 
More recently, a number of ULXs have been found to be supersoft; their 
X-ray spectra can be fitted with a blackbody model with
temperature of less than 100 eV (e.g., Swartz et al. 2002; Mukai et
al. 2003; Fabbiano et al. 2003; Di\,Stefano \& Kong 2003; Kong \&
Di\,Stefano 2003). 

In this Letter, we report a series of \chandra\ observations of the 
ultraluminous SSS in M101 (\sss) from the low state (Kong \& Di\,Stefano
2004) to outburst (Kong 2004a, 2004b). Shortly after 
the discovery of the outburst, we initiated a multiwavelength campaign to 
monitor the source.  We here also 
present the results of an \xmm\ follow-up observation conducted about two weeks 
after the first outburst observation. The primary result of radio 
observations by the Very Large Array was reported by Rupen, 
Sjouwerman, \& Kong (2004). We will present the detailed analysis of the 
optical and radio observations in a follow-up paper.

\section{Observations and Data Reduction}

\subsection{\chandra}

M101 was observed by \chandra\ in 2004 January, March, May, and 
July. Table 1 summarizes the \chandra\ observations. We also list 
the two \chandra\ observations taken in 2000 March and 2000 October, and 
the follow-up \xmm\ observation (see \S\,2.2). 
In all observations, the center of M101 and the ultraluminous SSS are in the S3 
chip of the ACIS-S. All data were taken in very faint mode (VFAINT). We 
reprocessed the raw data to utilize the VFAINT mode and 
to remove the 0.5-pixel randomization. 
However, for data taken during the outburst, we found that VFAINT 
processing flagged some source events of the SSS as background and we 
therefore did not apply VFAINT 
processing for the data taken in July 
\footnote{http://asc.harvard.edu/ciao/threads/aciscleanvf/}.
In order to reduce 
the instrumental background, we screened the data to allow only photon 
energies in the range of 0.3--7 keV. We also searched for periods of high
background using source free regions in the S1 chip and filtered all the 
high background periods. Only the 2004 January 24 and 2004 May 7 observations 
were affected; the exposure times listed in Table 1 correspond to the good 
time intervals. All data were reduced and analyzed with the
CIAO v3.1 package, and calibration database CALDB 2.27.

\subsection{\xmm}

Shortly after the discovery of the outburst of \sss\ with 
\chandra, we requested \xmm\ director's discretionary time for a follow-up 
observation.
The \xmm\ observation was taken on 2004 July 23 for 30 ks, and  
the instrument modes were full-frame, medium filter for the three 
European Photon Imaging Cameras (EPIC).
After rejecting those intervals with a high
background level, we considered a good time interval of $23$ ks. Only
data in 0.2--12 keV were used for analysis. Data were reduced and analyzed
with the \xmm\ SAS package v5.4.1.
%\footnote{http://xmm.vilspa.esa.es/external/xmm\_sw\_cal/sas\_frame.shtml}.

\section{Analysis and Results}

We performed spectral analysis for all \chandra\ data taken between
2004 January and 2004 July. 
The two datasets taken in 2000 were reported in Mukai 
et al. (2003) and Di\,Stefano \& Kong (2003). Spectral analysis was 
performed by using Sherpa. We
also used XSPEC v11.3 for independent check. 
In each observation, we extracted the source spectrum from a $6''$ radius
circular region 
centered on the source, while an annulus region centered on the source 
was used as the background.
During the outburst, there are enough photons to rebin the 
spectra with at least 15 counts per spectral bin, and used $\chi^2$ 
statistics to find the best-fitting parameters. For the data during the low 
state (January, March, and May), very few source photons were received 
($\sim 10-40$) in each observation, we therefore performed spectral analysis with 
unbinned data (background not subtracted) for which C-statistics (Cash
1979) were
employed. To examine the low state spectrum in detail,
we also combined all the low state data to perform spectral 
fit with binned data and $\chi^2$ statistics. We tried power-law, 
blackbody, and disk blackbody models with interstellar absorption.
Table 2 and 3 show the best-fitting spectral parameters for all \chandra\ 
observations taken in 2004, and the X-ray spectra are shown in Figure 1. 

\begin{inlinetable}
\caption{Observation Log of M101}
%\tiny
\begin{tabular}{lcc|lcc}
\hline \hline
\multicolumn{1}{c}{Date} & OBSID & Exposure & \multicolumn{1}{c}{Date} & OBSID & Exposure\\
\hline
2000-03-26 & 934 & 94 ks &  2004-05-07 & 4733 & 16 ks\\
2000-10-29 & 2065 & 9.6 ks & 2004-05-09 & 5323 & 43 ks\\
2004-01-19 & 4731 & 56 ks & 2004-07-05 & 5337 & 10 ks\\
2004-01-24 & 5297 & 15 ks & 2004-07-06 & 5338 & 29 ks \\
2004-03-07 &  5300 & 52 ks & 2004-07-07 & 5339 & 14 ks\\
2004-03-14 & 5309 & 71 ks & 2004-07-08 & 5340 & 54 ks\\
2004-03-19 & 4732 & 70 ks & 2004-07-11 & 4734 & 35 ks\\
2004-05-03 & 5322 & 65 ks & 2004-07-23 & 0164560701& 23 ks\\
\hline
\end{tabular}
\end{inlinetable}

For the low state data, power-law models provide better fits
than blackbody models. Each low-state observation is
consistent with an absorbed power-law model; the photon index ranges from 
$\sim 0.8-2.6$ among the observations, with an average of 1.8. The $N_H$ converged to zero and
we therefore fixed it at the Galactic 
value ($1.2\times10^{20}$ cm$^{-2}$; Dickey \& Lockman 1990). The $0.3-7$ keV 
luminosity is about $(1.5-4.6)\times10^{37}$\lum. Fixing the $N_H$ at
a higher value ($1.5\times10^{21}$ cm$^{-2}$; see below), the
luminosity is about a factor of 2 higher. The combined
low state data, however, cannot be fitted with any single-component
model. A simple power-law model produces a soft excess. We therefore
fitted the spectrum with a composite blackbody and power-law model
($N_H=1.5\times10^{21}$ cm$^{-2}$, $kT=63$ eV, and $\alpha=1.4$; see
Table 2); the power-law component contributes roughly $70\%$ (30\%) of
the absorbed (unabsorbed) 0.3--7 keV flux. 

During the 2004 outburst, blackbody models provide good fits to the spectra. Disk blackbody model is also equally acceptable; the inner disk
temperatures
are slightly higher than the blackbody temperatures, but the normalizations
(hence the inner disk radii) often have large error ranges. In general, the 
blackbody temperatures are between $\sim 50-100$ eV, and the $N_H$ is 
about $(1-4)\times10^{21}$ cm$^{-2}$, significantly higher than the
Galactic value. We also fitted the spectra with the $N_H$ fixed at
$1.2\times10^{20}$ cm$^{-2}$. However, all the fits are unacceptable with 
reduced $\chi^2$ between 1.5 and 2.4. 
The 0.3--7 keV luminosity ranges from $10^{39}$\lum\ to $10^{40}$\lum. We 
also derived the bolometric luminosity from the normalization of the 
blackbody model and it is between $5\times10^{39}$\lum\ and 
$10^{41}$\lum. In two cases, the spectra are complex and single
component models cannot provide 
good fits to the data. We included two absorption edges in each case and 
the fits were significantly improved. Specifically, we found absorption
edges at 0.66 keV and 0.87 keV on 2004 July 6 but the edges were not
observed on 2004 July 8. Instead, we found two new edges at 0.33 keV and
0.56 keV. 

We used similar techniques to analyze the \xmm\ data. We extracted the
source spectrum from a $15''$ radius circle centered on the source.
The source was
fainter than previous \chandra\ observations with about 
160 background-subtracted counts in EPIC-pn. 
Since the source was detected only in EPIC-pn, we fitted the EPIC-pn
spectrum with
several single-component models. However, all the fits are unacceptable. 
We therefore fitted the spectrum with a composite blackbody and 
power-law model. The model consists of a supersoft component (53 eV) and a 
hard power-law tail ($\alpha=0.7$); the power-law component
contributes about 40\% (8\%) of the 0.3--7 keV absorbed (unabsorbed) flux.
The 0.3--7 keV luminosity is 
$6.1\times10^{38}$\lum, indicating that the source was in the decay stage 
of the outburst.  

\begin{inlinefigure}
\epsfig{file=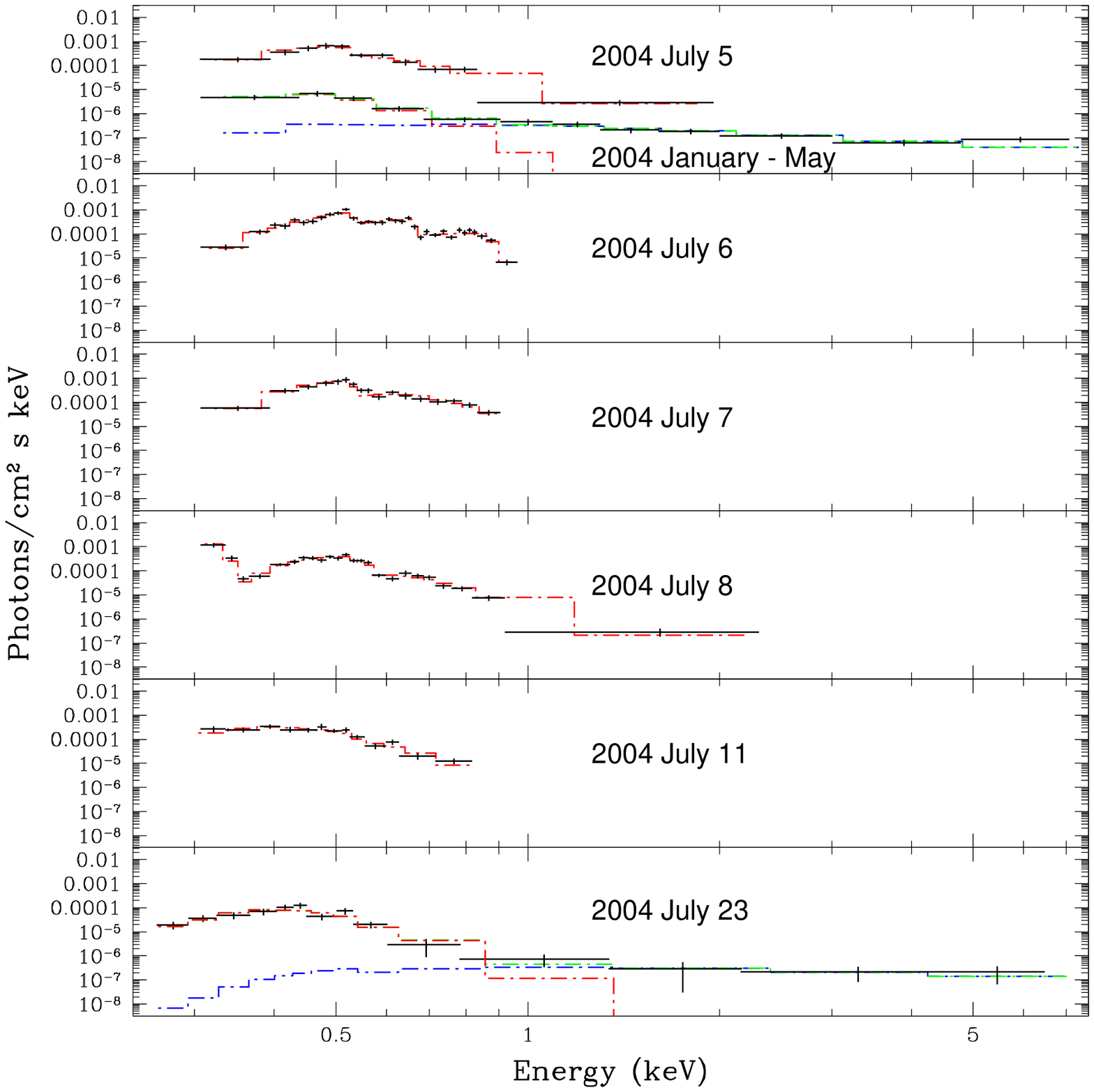,width=3.56in,height=3.9in}
\caption{Unfolded spectra of the ultraluminous SSS. The models
  correspond to the fits listed in Table 2 and 3. 
The total spectrum, blackbody component, and
  power-law component are shown in green, red, and blue, respectively.}
\end{inlinefigure}

\begin{table*}
\centering{
\small
\caption{Best-fitting Spectral Parameters for the Low State}
\begin{tabular}{lcccccccccc}
\hline
\hline
Date & \multicolumn{4}{c}{Power-law} & & \multicolumn{4}{c}{Blackbody} &\\
    \cline{2-5} \cline{7-10}\\ 
     & $N_H$$^a$ & $\alpha$ & MC$^b$ & $L_{0.3-7}$$^c$ & &$N_H$ &
$kT$ (keV) & $\chi^2_{\nu}/$dof, MC$^b$ & $L_{0.3-7}$$^c$ & $L_{Pow+BB}$$^d$\\
\hline
Jan 19 & 0.12$^e$ & $2.64^{+0.62}_{-0.58}$ & 0.73 & 0.15 && 0.12$^e$ &
$0.29^{+0.11}_{-0.18}$ & 1.0 & 0.10 & 0.71\\
Jan 24 & 0.12$^e$ & $1.46^{+0.83}_{-0.80}$ & 0.58 & 0.32 && 0.12$^e$ &
$0.66^{+0.41}_{-0.21}$ & 0.65 & 0.27 & 0.79\\
Mar 7  & 0.12$^e$ & $2.08^{+0.45}_{-0.43}$ & 0.81 & 0.27 && 0.12$^e$ &
$0.58^{+0.15}_{-0.10}$ & 0.99 & 0.25 & 1.04\\
Mar 14 & 0.12$^e$ & $1.94^{+0.48}_{-0.47}$ & 0.80 & 0.17 && 0.12$^e$ &
$0.63^{+0.19}_{-0.12}$ & 0.98 & 0.16 & 0.57 \\
Mar 19 & 0.12$^e$ & $1.89^{+0.45}_{-0.44}$ & 0.50 & 0.20 && 0.12$^e$ &
$0.41^{+0.16}_{-0.09}$ & 0.95 & 0.13 & 0.51\\
May 3  & 0.12$^e$ & $1.98^{+0.56}_{-0.53}$ & 0.52 & 0.15 && 0.12$^e$ &
$0.52^{+0.18}_{-0.11}$ & 0.82 & 0.12 & 0.49\\
May 7  & 0.12$^e$ & $0.78^{+0.77}_{-0.78}$ & 0.63 & 0.47 && 0.12$^e$ &
$1.51^{+3.60}_{-0.59}$ & 0.68 & 0.58 & 0.66 \\
May 9  & 0.12$^e$ & $1.54^{+0.63}_{-0.61}$ & 0.70 & 0.18 && 0.12$^e$ &
$1.00^{+0.61}_{-0.28}$ & 0.81 & 0.24 & 0.54\\
Jan--May$^f$ & $1.5^{+0.2}_{-0.9}$ & $1.41^{+0.29}_{-0.36}$ & --- &---
&& ---& $0.063^{+0.010}_{-0.015}$ & 1.05/7 &--- & 0.68 \\
Jul 23 & $1.5^{+1.9}_{-1.0}$ & $0.72^{+1.68}_{-1.98}$ & --- & --- &&
--- & $0.053^{+0.019}_{-0.016}$ & 1.12/9 &--- & 6.10 \\
\hline
\end{tabular}
}
\par
\medskip
%\hspace{0.005cm}
\begin{minipage}{0.94\linewidth}
\footnotesize
NOTES --- All quoted uncertainties are 90\% confidence. Sources
without reduced $\chi^2$ and dof values were fitted with unbinned data
using CASH statistics; ``goodness-of-fit''
determined by Monte-Carlo simulations (MC) is employed and parameters of
both blackbody and power-law models are shown. \\
$^a$ in units of $10^{21}$ cm$^{-2}$.\\
$^b$ For low state data, we list the probability that the best fit model
would produce a lower value of the CASH statistics than that calculated
from the data, as determined via XSPEC Monte-Carlo simulations.  A best
fit model should have a value of about 0.5. \\
$^c$ 0.3--7 keV luminosity ($\times 10^{38}$\lum), assuming 6.7 Mpc
(Freedman et al. 2001). \\
$^d$ 0.3--7 keV luminosity ($\times 10^{38}$\lum) derived from the combined power-law and
blackbody model.\\
$^e$ fixed\\
$^f$ Combined low-state data modeled by an absorbed power-law and
blackbody model.

\end{minipage}
\end{table*}

\section{Discussion}

Although over $100$ ULXs have been discovered (Miller \& Colbert 2003), only a handful of sources have $L_X\gaeq 10^{41}$ ergs s$^{-1}$, assuming isotropic radiation (Matsumoto et al. 2001; Gao et al. 2003; Davis \& Mushotzky 2004; Soria \& Motch 2004).
A luminosity this high is difficult to achieve in
an X-ray binary unless the accretor has a mass greater than
roughly $10\, M_\odot.$ We do have evidence that \sss\
is likely to be an X-ray binary, since its luminosity has been
observed to change by a factor of $\gaeq 10$ on a time scale of hours (Mukai et al. 2003; Di\,Stefano \& Kong 2003).
While it is possible that our luminosity estimates are 
higher than the true luminosity, the effects that lead to
overestimates, such as beaming, or various anisotropies,
tend to change estimated luminosities downward by a factor of roughly  
$10.$ Its unusually high bolometric luminosity ($1.2\times 10^{41}$ ergs
s$^{-1}$), coupled with its
short-time-scale time variability, therefore make \sss\
a good candidate for an accreting IMBH.  For instance, King et
al. (2001) 
proposed that anisotropic
X-ray emission can result super-Eddington luminosity for a
stellar-mass BH. However, in order to achieve such a high luminosity
for a stellar-mass BH,
extreme beaming is required and the disk is expected to be
much hotter. Similarly, the temperature of radiation pressure-dominated accretion disk
model proposed by Begelman (2002) is too high.
The pure blackbody spectrum
also makes relativistically beaming (K{\"o}rding et
al. 2002) unlikely. 

Its soft spectrum in the high state is another important piece of evidence.
%Such a soft spectrum is consistent with SSSs seen in
%our own Galaxy, Magellanic Clouds (Greiner 2000 \footnote{see
%http://www.mpe.mpg.de/$\sim$jcg/sss/ssscat.html for updated catalog}) and
%several nearby galaxies (e.g., Swartz et al. 2002; Di\,Stefano \& Kong
%2003; Di\,Stefano et al. 2004).
An accretion disk around a very massive
BH is expected to produce supersoft X-ray emission (see Di\,Stefano \&
Kong 2002). For instance, if the supersoft emission comes from the
inner disk, a 70 eV SSS at luminosity of $10^{40}$ \lum\
corresponds to a $\sim 6000~ M_{\odot}$ BH. If we used the 90\% lower
limits of the inner disk temperature derived from the disk blackbody fits,
the BH mass is estimated to be $> 2800 M_\odot$ (Makishima et al. 2000).
This is consistent with 
prior work on IMBH models for 
ULXs (see e.g., Miller \& Colbert 2003). It also complements other
work on evidence for cool disks
in ULXs that has also been considered as evidence for 
IMBH models (Miller et al. 2003,2004) although a power-law component
usually contributes a significant fraction of X-ray emission. Furthermore,
the high state luminosity is approximately
$16\%$ of the Eddington luminosity for a $6000\, M_\odot$ BH;
we therefore expect the inner disk to be optically thick, which is consistent
with the IMBH interpretation.   

\begin{table*}
\centering{
\small
\caption{Best-fitting Blackbody Parameters for the High State}
\begin{tabular}{lcccccccccc}
\hline
\hline
Date & $N_H$ & $kT$ & E$_{edge 1}$ & $\tau_1$&
E$_{edge 2}$ & $\tau_2$ & $\chi^2_{\nu}/$dof &
$L_{0.3-7}$ & $L_{bol}$ \\
     & ($10^{21}$ cm$^{-2}$) & (eV) & (keV) & & (keV) & & & ($
10^{40}$\lum) & ($10^{40}$\lum)\\
\hline
Jul 5  & $2.2^{+0.2}_{-0.2}$ & $70^{+15}_{-15}$ &--- &--- &---&---&0.90/8 
& 0.74 & 1.97\\
Jul 6  & $2.9^{+0.2}_{-0.2}$ & $104^{+22}_{-22}$ &
$0.66^{+0.02}_{-0.02}$ &$1.6^{+0.6}_{-0.6}$ & $0.88^{+0.03}_{-0.03}$ &
$2.8^{+1.9}_{-1.9}$ & 0.97/26 & 1.04 & 1.82\\
Jul 7  & $3.8^{+2.8}_{-2.8}$ & $64^{+17}_{-17}$ & ---&--- &--- &--- & 0.98/14 & 3.82 & 12.40\\
Jul 8  & $1.4^{+0.6}_{-0.6}$ & $56^{+4}_{-4}$ & $0.33^{+0.05}_{-0.05}$
& $5.3^{+4.0}_{-4.0}$ & $0.57^{+0.03}_{-0.03}$ & $1.2^{+0.6}_{-0.6}$ &
1.19/17 & 0.81 & 8.71\\
Jul 11 & $1.2^{+1.3}_{-1.0}$ & $59^{+11}_{-10}$ & ---&---  &--- &--- & 1.26/10 & 0.17 & 0.63\\
\hline
\end{tabular}
}
\end{table*}

The state changes of \sss\ span a luminosity range larger than a factor
of $1000.$ \sss\ 
was first detected by \rosat\ in 1996 November (Wang et al. 1999),
with an extrapolated unabsorbed 0.3--7 keV luminosity of $8\times10^{38}$ \lum\ (blackbody
model with $N_H=10^{21}$ cm$^{-2}$, $kT=75$ eV). Higher absorption
($3\times10^{21}$ cm$^{-2}$) would imply a luminosity up of
$4\times10^{39}$\lum. The source was not detected in other \rosat\
observations (see Wang et al. 1999). In a 108 ks \rosat\ HRI
observation taken in 1996 May/June, we can set a
$3\sigma$ detection limit at $6.8\times10^{37}$\lum\ (power-law model
with $N_H=1.2\times10^{20}$ cm$^{-2}$ and $\alpha=2$). Setting the
$N_H$ at $1.5\times10^{21}$ cm$^{-2}$, the luminosity limit becomes
$1.5\times10^{38}$\lum. In 2000 March and October, \chandra\ detected
the SSS in the very soft state with luminosities of $\sim 10^{39}$\lum\ (Mukai et al. 2003;
Di\,Stefano \& Kong 2003).
In 2002 June, \xmm\ observed M101 and did not
detect the SSS (Jenkins et al. 2004). We re-analysed the image and found that the SSS
may be barely detected with possible contamination from nearby sources. We
derived the 0.3--7 keV luminosity of the SSS as
$\sim 10^{37}$\lum, by assuming a power-law model. The source was also in
the low state between 2004 January and May with luminosity of
$\sim10^{37}$\lum. It is clear that
the SSS has had at least three major outbursts ($L_X > 10^{39}$\lum). On the other hand, the
very low luminosities ($\sim 10^{37}$\lum)
during 2002 and 2004 indicate that the source was in the low state.
The source varies by as much as a factor of 1000 between the
low state and the high state. This amplitude is even greater than many
Galactic BHs. Remarkably, the source also shows spectral changes.
It is very clear that there is a power-law
component ($\alpha=1.4$) in the composite low state spectrum, while the
high state spectra are supersoft. 
Our \xmm\ observation taken 
during the decay of the outburst revealed that the blackbody
component was still strong and there was a very hard power-law
tail, similar to the low state. While it is difficult to make direct
comparison with Galactic BHs because of the blackbody emission in
the low state and the pure blackbody spectrum in the high state (see
however Zdziarski \& Gierlinski 2004), it is unusual
for a SSS to have high/very soft state and low/hard state
transition. 
With a peak bolometric luminosity $\approx 10^{41}$\lum, it is
likely that IMBH is the central engine of the system. Interestingly,
the blackbody component is always seen while the power-law component
becomes stronger when the source is at lower
luminosities. It may indicate that the power-law component is due to
Comptonization of soft photons (Wang et al. 2004). The photon index is,
however, harder
than that of Galactic BHs in the low state ($\alpha\sim1.7$). None of
the Galactic BHs has similar spectrum (McClintock \& Remillard
2003). The closest example is Galactic microquasar V4641 Sgr for which the
photon index was measured to be between 0.6 and 1.3 (Miller et
al. 2002; McClintock \& Remillard 2003). 

Another important feature of our X-ray spectral fits is the presence
of absorption edges, which we found at 0.33,
0.57, 0.66, and 0.88 keV in two of the high state spectra. These are
consistent with C {\small V}, N {\small VI}, N {\small VII}, and O {\small VIII} edges. We note, however,
that the 0.33 keV edge may be due to calibration of the ACIS-S near
the carbon edge around 0.28 keV. 
These features may signal the presence of highly ionized
gas in the vicinity of the accretor (e.g., warm absorber), and may be consistent
with an outflow from the source.  
In fact, outflow models
have been suggested for ULXs. King and Pounds (2003) suggested that the soft
component of some ULXs is due to outflow of a stellar-mass BH. Mukai
et al. (2003) also used similar argument to explain the
super-Eddington luminosity of \sss. However, the new data which reveal the changes of
temperature and bolometric luminosity, the extremely high
luminosity, and the state transition of the source will be difficult
to explain by such a model. It
remains a puzzle that we saw different edges in the two observations,
possibly related to the geometry of the system.
Nevertheless, outflows from IMBHs may be expected. 

It is also interesting to note that similar absorption edges 
are expected and have been
observed in white dwarf (WDs) systems such as the recurrent nova 
U Sco (Kahabka et al. 1999), CAL~87 and RX~J0925.7--4758 (Ebisawa et
al. 2001). SSS radiation  with
luminosities in the range from $10^{37}$ ergs s$^{-1}$ to
roughly the Eddington limit for a $1.4\, M_\odot$ WD are
expected for nuclear burning WDs (van den Heuvel et al. 1992). Indeed,
such models have been proposed to explain SSSs in the
Galaxy and Magellanic Clouds. Note, however, that neutron  star and
BH models have also been proposed. The high state luminosity of this source
is almost two orders of magnitude larger than that of any other
known SSS, and it rules out steady nuclear burning on hot WD models as an explanation for \sss .
A similar argument can be made for neutron star models.  
A nova explosion could explain some features of the data. It is not favored, however, because the photospheric radius and lack of hard radiation at the peak of the outburst have no obvious explanation.

It may be impossible to conclusively establish, using current technology, that
any X-ray source in an external galaxy is an IMBH.
A system like \sss\ is therefore particularly valuable, because
it provides $4$ different pieces of evidence that together
make a consistent argument in favor of an IMBH interpretation.  
Its high-state luminosity, its short-time-scale
variability, its soft high-state spectrum, and its pronounced
spectral changes, all suggest that \sss\ is a strong IMBH candidate.

\begin{acknowledgements}
This work was supported by NASA grant NNG04GP58G.
A.K.H.K. acknowledges support from NASA grant GO3-4049X from the 
Chandra X-Ray Center, and would like to 
thank \xmm\ project scientist
Norbert Schartel for granting our DDT request. 
F. Y. acknowledges support from NASA grant NAG5-9998.
The 2004 \chandra\ data are part of the Very Large Project to study M101 (PI: K. Kuntz). We thank R. Soria for useful discussion.
This work is based on observations obtained with \xmm, an ESA mission with
instruments and contributions directly funded by ESA member states and
the US (NASA). 
\end{acknowledgements}

\end{document}